\documentclass[a4paper,11pt,twoside]{article}
\usepackage{amssymb,amsmath,amsthm,verbatim,natbib}
\usepackage[dvips]{graphicx}
\usepackage{setspace}
\usepackage{fancyhdr}

\def\P{\mathop{\rm P}\nolimits}%
\def\U{\mathop{\rm U}\nolimits}%
\def\N{\mathop{\rm Normal}\nolimits}%
\def\E{\mathop{\rm Exp}\nolimits}%
\def\G{\mathop{\rm Gamma}\nolimits}%

\newtheorem{proposition}{Proposition}
\newtheorem{theorem}{Theorem}
\newtheorem{remark}{Remark}

\begin{document}

\vspace*{0.05cm}

\begin{flushleft}
{\huge \bf A Statistical Analysis of Probabilistic Counting Algorithms} \\

\vspace{0.4cm}

{\large PETER CLIFFORD \\
Department of Statistics, University of Oxford \\

\vspace{0.2cm}

IOANA A.\ COSMA \\
Statistical Laboratory, University of Cambridge} 
\end{flushleft}

\vspace{0.2cm}

\noindent{\bf ABSTRACT.  This paper considers the problem of cardinality estimation in data stream applications.  We present  a statistical analysis of probabilistic counting algorithms, focusing on two techniques that use pseudo-random variates to form low-dimensional data sketches. We apply conventional statistical methods to compare probabilistic algorithms based on storing either selected order statistics, or random projections. We derive estimators of the cardinality in both cases, and show that the maximal-term estimator is recursively computable and has exponentially decreasing error bounds.  Furthermore, we show that the estimators have comparable asymptotic efficiency, and explain this result by demonstrating an unexpected connection between the two approaches.}

\vspace{0.2cm}


\section{Introduction}

High-throughput, transiently observed, data streams pose novel and challenging problems for computer scientists and statisticians \citep{Muthukrishnan.05, Aggarwal.07}.  Advances in science and technology are continually expanding both the size of data sets available for analysis and the rate of data acquisition; examples include increasingly heavy Internet traffic on routers \citep{ABRS.03, CM.04b}, high frequency financial transactions, and commercial database applications \citep{WVT.90}.  

The online approximation of properties of data streams, such as cardinality, frequency moments, quantiles, and empirical entropy, is of great interest \citep{CM.04a, HNO.08.2}.  The goal is to construct and maintain sub-linear representations of the data from which target properties can be inferred with high efficiency \citep{Aggarwal.07}. Data stream algorithms typically allow only one pass over the data, i.e., data are observed, processed to update the representation, and then discarded.  By `efficient' with respect to the inference procedure, we mean that estimators are accurate with high probability.  With respect to the handling of data, we mean that the algorithm has fast processing and updating time per data element, uses low storage, and is insensitive to the order of arrival of data.  This is in contrast to sampling-based techniques, that are sensitive to the pattern of repetitions in the data.  

This article focuses on the problem of estimating the number of distinct items in a data stream when storage constraints preclude the possibility of maintaining a comprehensive list of previously observed items.  The number of distinct items or {\em cardinality} can, for example, refer to pairs of source-destination IP addresses, observed within a given time window of Internet traffic, monitored for the purpose of anomaly detection, e.g., denial-of-service attacks on the network \citep{Giroire.09}.  There is a surprisingly long history of work on cardinality estimation in the computer science literature, starting from the pioneering work of \cite{FM.85}, and developed in isolation from mainstream statistical research. Our purpose is to re-analyse these algorithms in `traditional' statistical terms.   The focus will be on comparisons in terms of asymptotic relative efficiency, pivotal quantities, and statistical errors bounds, as opposed to the focus in the computer science literature on storage space and processing time.  We will concentrate on {\em sketching algorithms} that exploit hash functions to record meaningful information, either by storing order statistics or by random projections.  

The paper is organised as follows. In Section~\ref{sec:recentWork} we define terms, such as hash function and hashing, and give a brief and selective history of cardinality estimation algorithms.  We then investigate two types of algorithms from a conventional statistical viewpoint, deriving maximum likelihood estimators (MLEs) for methods based on order statistics in Section~\ref{sec:orderstats} and random projections in Section~\ref{sec:randomproj}. For order statistic methods, we show that the choice of sampling distribution is immaterial when sampling from a continuous distribution, but that substantial savings in storage can be achieved by using samples from the geometric distribution without significant reduction in the asymptotic relative efficiency. We also show that these estimators are recursively computable with exponentially decreasing error bounds. We then propose an approximate estimator for projection methods using $\alpha$-stable distributions, with $\alpha$ close to zero.  In Section~\ref{sec:compare} we compare the two methods and find unexpectedly that, in a certain sense, they are essentially equivalent. Finally, in Section~\ref{sec:simulations}, we compare the performance of our algorithms to existing benchmark algorithms on simulated data.  Section~\ref{sec:conclude} concludes the paper.

\section{Definitions and history} \label{sec:recentWork}
 
We define a {\em discrete data stream} to be a transiently observed sequence of data elements with types drawn from a countable, possibly infinite, set $\mathcal{I}$.  At discrete time points $t=1,\dots,T$, a pair of the form $(i_t,d_t)$ is observed, where $i_t \in \mathcal{I}$ is the type of the data element, and $d_t$ is an integer-valued quantity.  Let $\mathcal{I}_T$ be the set of distinct data types observed by time $T$. 
 
A basic goal in data stream analysis is to obtain information about the collection $\textbf{a}(T)=\left\{a_i(T),i \in \mathcal{I}_T \right\}$, where  $a_i(T) = \sum_{t=1}^T d_t \mathbb{I}(i_t = i)$ is the cumulative quantity of type $i$ at time $T$.  When there is no possibility of confusion, we write $\textbf{a}$ and $a_i$ for $\textbf{a}(T)$ and $a_i(T)$, respectively.  Our concern will be primarily with the special case when $d_t > 0, \forall t$; the {\em cash register} case in the terminology of \cite{CDIM.02}. Many summary statistics of interest are functions of $\textbf{a}$, e.g., $c = \sum_{i \in \mathcal{I}_T  } \mathbb{I}(a_i(T)  >  0)$, the cardinality of the set $\mathcal{I}_T$ in the cash register case. Recall that we are assuming that storage constraints make it impossible to know $\mathbf{a}$ precisely.
 
{\em Hashing} \citep{Knuth.98} is a basic tool used in processing data, where the type of data element is identified by a complicated label.  Hashing was originally designed to speed-up table lookup for the purpose of item retrieval or for identifying similar items.  For example, suppose that data elements are records of company employees, uniquely identified by complicated labels, that must be stored in a table.  A {\em hash function} can be designed to map the label to an integer value in a given range, called the {\em hash value}, indexing the location in the table where the corresponding employee record is stored.  \cite{PTVF.07} present algorithms for constructing hash functions.  Given the hash function and a label, the corresponding record is easily accessible for updating, for example.   

In general, a hash function $h: \mathcal{I} \mapsto \{1,\ldots,L\}$ is a deterministic function of the input in $\mathcal{I}$ that has low collision probability, i.e., $\P \big( h(i) = h(j), i \neq j \big ) < 1/L$, where a collision occurs if two or more different inputs are mapped to the same hash value \citep{Knuth.98}.   A truly random hash function maps values in $\mathcal{I}$ to $\{1,\ldots,L\}$ independently; however, no construction exists for such functions.  Instead, the requirement for independence is reduced to $k$-wise independence, where any $k$ distinct values in $\mathcal{I}$ are mapped to $k$ independent values in $\{1,\ldots,L\}$.  \cite{CW.79} is the first reference on constructing $k$-wise independent hash functions. 

For our purposes, we can think of a hash function as the mapping between the seed of a random number generator and the first element in the sequence of computer generated pseudo-random numbers, usually uniformly distributed over some range. A collection of independent hash functions $(h_1,\dots,h_m)$ then corresponds to the $m$ individual mappings from the seed to the first $m$ elements of a pseudo-random sequence. This method of constructing a hash function mapping to pseudo-random numbers having a given distribution is known as the {\it method of seeding}.   \cite{Nisan.92} shows that there exists an explicit implementation of a pseudo-random number generator that converts a random seed, i.e., in our case, an element in $\mathcal{I}$, to a sequence of bits, indistinguishable from truly random bits.  Hence, we can assume throughout that the sequences underlying our hash functions are truly random.

\subsection{Probabilistic counting}
\cite{FM.85} introduce the idea of independently hashing each element $i \in \mathcal{I}_T$ to a long string of pseudo-random bits, uniformly distributed over a finite range.  Let $\rho(i)$ denote the rank of the first bit 1 in $h(i)$.  The algorithm stores and updates a bitmap table of all the values of $\rho$ observed, and returns an asymptotically unbiased estimate of the cardinality based on the quantity $\max \big \{ r; [1,\ldots,r] \subseteq \{\rho(i), i \in \mathcal{I}_T \} \big \}$.  The LogLog counting algorithm \citep{DF.03} estimates the cardinality from the summary statistic $\max_{i \in \mathcal{I}_T} \rho(i)$, avoiding the need for the bitmap table.  The algorithm offers an improvement in terms of storage requirements, for given accuracy, by storing small bytes rather than integers.   The Hyper-LogLog counting algorithm \citep{FFGM.07} improves the accuracy further by proposing a harmonic mean estimator based on this maximum statistic.  The latter algorithm is particularly well suited for large scale cardinality estimation problems.  \cite{CC.09} develop an algorithm that combines hashing to bit patterns with sampling at an adaptive rate.  They show empirically that their algorithm outperforms Hyper-LogLog for small to medium scale problems, but lack theoretical justification of this claim.

Instead of estimating the cardinality from bit patterns, \cite{Giroire.09} hashes the data types uniformly to pseudo-random variables in (0,1), stores order statistics of hash values falling in disjoint subintervals covering this range, and averages cardinality estimates over these subintervals.  This approach is called {\em stochastic averaging} and was introduced by \cite{FM.85}.    The MinCount algorithm \citep{Giroire.09} stores the third order statistic, and employs a logarithmic family transformation.  Table~\ref{table:compare} shows that these estimators have comparable precision.  The asymptotic relative efficiency (ARE) is defined as the ratio of $c^2/m$ to the asymptotic variance of the estimator.  \cite{CG.06} show that, in a large class of nearly-unbiased cardinality estimators based on order statistics, the variance of an estimator is lower-bounded approximately by $c^2/m$, where stochastic averaging over $m$ intervals is employed.  This equals the asymptotic variance of our estimator based on order statistics from Section~\ref{sec:orderstats}.

\begin{table}[!ht] 
\caption{A comparison of cardinality estimation algorithms based on hashing and storing order statistics, or random projections.  The first four algorithms apply stochastic averaging with $m$ subintervals.  Float stands for floating point number.}\label{table:compare}
\begin{center}
\begin{tabular}{|c|c|c|}
\hline
Algorithm & Cost & ARE \\
\hline
Probabilistic counting \citep{FM.85} & $m$ integers (16-32 bits) & 1.64 \\
LogLog \citep{DF.03} & $m$ small bytes (5 bits) & 0.592 \\
Hyper-LogLog \citep{FFGM.07} & $m$ small bytes & 0.925 \\
MinCount \citep{Giroire.09} & $m$ floats (32-64 bits) & 1.00 \\
Maximal-term (continuous) & $m$ floats & 1.00 \\
Maximal-term (geometric, $q=1/2$) & $m$ integers & 0.930 \\
Maximal-term(geometric, $q=10/11$) & $m$ integers & 0.999 \\
Random projections (Proposition~\ref{thm:projection}) & $m$ floats & 1.00 \\
Random projections \citep{CDIM.02} & $m$ floats & 0.481 \\
\hline
\end{tabular}
\end{center}
\end{table}

Projection methods for $l_{\alpha}$ norm estimation with streaming data are described in \cite{Indyk.06} for $\alpha \in \{1,2\}$,  and references therein.  The idea is to hash distinct data types $i_t$ to independent copies of $\alpha$-stable random variables, and store weighted linear combinations of the hash values.  Exploiting properties of the stable law, \cite{CDIM.02} approximate the cardinality using estimates of $l_{\alpha}$ with $\alpha$ close to zero.  The seminal paper of \cite{AMS.99} is the first attempt at obtaining tight lower bounds on the space complexity of approximating the cardinality of a simple data stream.   \cite{BYJKST.02} present the best previous $(\epsilon,\delta)$-approximation of the cardinality of a simple data stream in terms of space requirements, namely $O \left ( 1/\epsilon^2 \cdot \log(\log c) \cdot \log(1/\delta) \right )$; this work is the first to make no assumptions on the existence of a truly random hash function.  An estimator $\hat{c}$ is said to be an {\it $(\epsilon, \delta)$-approximation} of $c$, for some $\epsilon, \delta >0$ arbitrarily small, if $\P \left (|\hat{c}-c| > \epsilon c \right ) \leq \delta$. \cite{IW.03} show that the dependence of the space requirement on $\epsilon$ through the factor $1/\epsilon^2$ cannot be reduced to $1/\epsilon$.   \cite{KNW.10} offer the best algorithm for an $(\epsilon, \delta)$- approximation of the cardinality of a simple data stream with space requirement of $O \left (1/\epsilon^2 + \log(c) \right)$, and no assumptions on the existence of a truly random hash function.  For a general data stream, the $(\epsilon, \delta)$-approximation of \cite{CDIM.02} requires a data sketch of length $O \left (1/\epsilon^2 \cdot \log(1/\delta) \right )$; this result is obtained from Chernoff bounds on tail probabilities of the estimator $\hat{c}$. We employ the same approach in Section~\ref{sec:storage} to derive storage requirements for our algorithms.  

\section{Order statistics} \label{sec:orderstats}
\subsection{Continuous random variables}\label{sec:cont} 

A data stream in the cash register case provides data elements of the form $(i_t,d_t)$, where $i_t \in \mathcal{I}_T$, and $d_t > 0$, for $t=1,\ldots,T$. We start with a simple adaptation of the ideas of \cite{FM.85} and \cite{Giroire.09}, which we call the {\em maximal-term data sketch}.  At time $t$, the data type $i_t$ is used as the seed of a random number generator to produce the first pseudo-random number $h(i_t)$ uniformly distributed on (0,1).  Write $h(i_t) \sim \U(0,1)$.  The algorithm records $h^+$, the maximum value of $h(i_t)$, as the stream is processed, restarting the random number generator with the seed $i_t$ at each stage. Note that if a particular data type is seen more than once, the value of $h^+$ is unchanged, but whenever a new type $i_t$ is observed, there is a chance that $h^+$ will increase.  

For the idealised $\U(0,1)$ hash function, the variable $Y=h^+$ has density $f(y;c) = c y^{c-1}, \; y \in (0,1),$ since it is the maximum of $c$ independent $\U(0,1)$ variables, where $c$ is the unknown cardinality.  The quantity $c$ is then an unknown parameter to be estimated by standard statistical methods.  To increase the efficiency in estimating $c$, we sample $m$ successive values $h_1(i_t),\dots,h_m(i_t)$ from the random number generator at each stage, and store $Y_j = h^+_j$, $j=1,\dots,m$, thus obtaining a sample of size $m$ from $f(y;c)$.  
\begin{proposition}\label{thm:maxterm}
The MLE of $c$ based on $(Y_1, \dots, Y_m)$ is $\hat{c} = -m / \sum_{j=1}^m \log Y_j$ with asymptotic distribution $\N(c, c^2/m)$ as $m \to \infty$. The expression 
$
- c \sum_{j=1}^m \log(Y_j) \sim \G(m,1)
$
can be used as a pivot in setting exact confidence intervals for $c$.  
\end{proposition}
\begin{proof}
Using standard sampling theory.
\end{proof}

Asymptotically, $\hat{c}$ is unbiased and approximately normally distributed with standard error  $\hat{c}/\sqrt{m}$, so that by storing $m = 10,000$ values, for example, we can obtain an estimate of $c$ to within $2\%$ with $95\%$ confidence, regardless of the size of $c$.  

\begin{remark} When estimating an integer valued parameter, such as the cardinality, the derivatives involved in the standard derivation of the large sample distribution of the MLE cannot be calculated.  Nevertheless, equivalent results can be derived in terms of finite differences, and since the standard deviation of the estimators we consider is of the order of $c$, with $c$ large, the use of derivatives can be justified.  \cite{Hammersley.50} provides an early discussion of these issues.
\end{remark}

Note that the maximal-term sketch does not allow deletions in the stream, i.e., $d_t < 0$, since it does not take into account the value of $d_t$, and thus cannot modify the quantities $Y_j$ if $a_{i_t}$ becomes zero.   In contrast, the method of data sketching via random projections in Section~\ref{sec:randomproj} allows deletions and permits the estimation of $\sum_{i \in \mathcal{I}_T} \mathbb{I}(a_i(T)  >  0)$, provided that $a_i(T) \geq 0$ whenever the estimation procedure is applied.
 
\subsubsection{Using the $k$th order statistic}
A possible improvement might be to store the $k$th order statistic of the hash values rather than $h^+$.
\begin{proposition}
For $k < c$, let $Y_j$ denote the $k$th order statistic of the hash values from the $j$th hash function $h_j \sim \U(0,1), j=1,\ldots,m$.  The MLE $\hat{c}$ of $c$ based on $Y_1 = y_1, \ldots, Y_m = y_m$ is the unique root of 
\begin{equation} \label{eq:mle} 
\log \left( \prod_{j=1}^m y_j \right) + \sum_{i=1}^k\frac{m}{\hat{c}-i+1}  = 0.
\end{equation} 
When $c$ is large, the root is given approximately by 
$ 
\hat{c} = k \left (1-\prod_{j=1}^m y_j^{1/m} \right )^{-1},
$
with standard error approximately $\hat{c}/\sqrt{km}$.  Furthermore, the estimator in \eqref{eq:mle} is recursively computable.
\end{proposition}
\begin{proof}
The first part of the proof is straightforward.  For the second, recall that a sequence of statistics $T_m(x_1,\ldots,x_m)$ is said to be {\it recursively computable} if  
\begin{displaymath}
T_m(x_1,\ldots,x_m) = T_m(z_1,\ldots,z_m) \Rightarrow T_{m+1}(x_1,\ldots,x_m,w) = T_{m+1}(z_1,\ldots,z_m,w), \forall m \in \mathbb{N};
\end{displaymath}
see for example \cite{Lauritzen.88} who proves, for independent random variables $X_1,\ldots,X_m$, that if $T_m(X_1,\ldots,X_m)$ is minimal sufficient, then the sequence $T_m$, $m \geq 1$ is recursively computable. This property of sufficient statistics was first remarked by \cite{Fisher.25}.  It follows from a theorem of \cite{LS.50} that the statistic $T_m(Y_1,\ldots,Y_m) = \prod_{j=1}^m Y_j$ is minimal sufficient for $c$, so $\hat{c}$ is also minimal sufficient and hence recursively computable.
\end{proof}

The property of recursive computability is particularly important when dealing with massive data sets due to constraints on available storage.  For example, suppose two independent estimates, $\hat{c}_{1}$ and $\hat{c}_{2}$, of the cardinality $c$ are available, based on samples of size $m_1$ and  $m_2$. By substituting the estimates in \eqref{eq:mle}, the associated product terms can be recovered; the combined estimate can then be obtained by combining the products and using \eqref{eq:mle} once again with $m=m_1+m_2$.  When $c$ is large, the combined estimate is approximated by   
\begin{displaymath} 
\frac{k}{1 - \left [\left (1- k / \hat{c}_{1} \right )^{m_1} \left (1- k / \hat{c}_{2} \right )^{m_2} \right ]^{1/(m_1 + m_2)}}.         
\end{displaymath}

Furthermore, we remark that to keep a record of the $k$th order statistic for each of the $m$ subsets as the stream is processed requires storing $km$ values.  However, since the standard error of $\hat{c}$ is approximately $\hat{c}/\sqrt{km}$ for large $m$, there is no gain in accuracy relative to the storage requirement.  

We also note that there is no advantage in using a hash function $h$ that maps to a continuous distribution $F$ other than $\U(0,1)$. The MLE of the maximal-term data sketch merely becomes 
\begin{equation}\label{eq:general_max} 
\hat{c} = -\frac{m}{\sum_{j=1}^m \log F(M_j)}, \quad \text{where} \ M_j = \max_{i \in \mathcal{I}_T}h_j(i),
\end{equation} 
which has the same distribution as $\hat{c}$ in Proposition~\ref{thm:maxterm}.  

\subsection{Discrete random variables}\label{sec:discrete}
Hashing to integer values rather than floating point numbers requires less storage, a priority when handling massive data streams.  We show that the loss of statistical efficiency is negligible when integer-valued hash functions are chosen appropriately.  We first consider hashing to Bernoulli random variables, not previously considered in the literature, and then to geometric random variables.

\subsubsection{Data sketching with Bernoulli random variables} 
To implement hashing to a Bernoulli variable, we start with an array of $0$s of length $m$ and then change the $j$th element to $1$ if $h_j(i_t) < p$, where, as before, $h_j(i_t)$ is the $j$th simulated $\U(0,1)$ variable from the seed $i_t$, $j=1,\dots,m$. The value of $p$ is chosen to maximise Fisher's information. 

\begin{proposition} \label{prop:MLE_Bernoulli} 
Fisher's information for a Bernoulli hash functions with probability $p$ is maximised with $p_{\rm max}= 1-\exp(-\lambda_0/c) \approx \lambda_0/c$, for large $c$, where $\lambda_0 = 2+\text{W}(-2 e^{-2}) \approx 1.594$, and $W$ is Lambert's function.   The asymptotic relative efficiency of the MLE of $c$ with Bernoulli hashing $(p=\lambda/c)$, relative to the estimator obtained with a continuous hash function, is $\lambda^2/(e^{\lambda} - 1)$ for large $c$.  
\end{proposition}    

This result enables lower bounds on the asymptotic relative efficiency to be specified.  For example, if $c$ is known in advance to lie in $(0.3c_0,4.3c_0)$ for some fixed $c_0$, then with $p = 1/c_0$ the ARE is at least $25\%$.  Consequently, $4m$ bits of storage suffice to provide the same accuracy as storing $m$ floating point numbers when hashing to continuous random variables. 

\begin{proof}[Proof] 
After processing the data stream we have observations from $m$ Bernoulli variables, each with probability $P=1-(1-p)^c$.  Fisher's information for $P$ is $m/(P(1-P))$ and hence the information for $c$ is 
\begin{displaymath}
I(c) = \frac{m}{P(1-P)} \left(\frac{dP}{dc}\right)^2 =  \frac{m q^c(\log q)^2}{1-q^c},
\end{displaymath}
where $q=1-p.$  Substituting $q = \exp(-\lambda/c)$ gives $ I(c) = m c^{-2}\lambda^2 /(e^\lambda -1)$. Since Fisher's information using continuous variables is $m/c^{2}$, this gives the asymptotic relative efficiency as claimed. The Fisher information from Bernoulli hashing attains its maximum when $\lambda$ is the positive root of $\lambda = 2(1-\exp(-\lambda))$, which can be expressed in terms of Lambert's $W$ function and is given approximately by $\lambda_0 = 1.594$.        
\end{proof}

\subsubsection{Geometric random variables} 
Suppose that the hash function maps to a geometric random variable with cumulative distribution function $G_p(x)=1-q^x$, with $p+q=1$, $x=1,2,\ldots$ We note that $p=1/2$ is the case analysed by \cite{DF.03} and \cite{FFGM.07}.  As before, for the maximal-term data sketch, we store $Y_j = h^+_j = \max \left \{h_j(i_t); i_t \in \mathcal{I}_T \right \}$, $j=1,\dots,m$, where $h_j(i_t)$ are independently simulated from $G_p$ by the method of seeding, and estimate $c$ based on the random sample $Y_1=y_1, \ldots, Y_m=y_m$.  Let $G_p^{\,c}$ be the distribution function of the maximum of $c$ independent $G_p$ variables.  

\begin{proposition} \label{prop:MLE_Geom}
The MLE of $c$ based on a sample $Y_1=y_1, \dots, Y_m=y_m$ drawn from $G_p^c$ satisfies 
\begin{equation} \label{eq:discrete} 
\sum_{i=1}^m \frac{\log\big(1-q^{y_i} \big) \big (1-q^{y_i}\big )^{\hat{c}} - \log \big(1-q^{y_i-1} \big)
\big(1-q^{y_i-1}\big)^{\hat{c}}}{\big(1-q^{y_i}\big)^{\hat{c}} - \big(1-q^{y_i-1}\big)^{\hat{c}}}=0. 
\end{equation} 
In the limit as $m \to \infty$, the distribution of $\hat{c}/c$ is asymptotically normal with mean $1$ and variance $1/(m\psi_c)$ where  $\psi_c$ can be approximated by
$$\psi_\infty = \sum_{k=-\infty}^{\infty}  \frac{q^{2k}(q^{-1}-1)^2}{[\exp(q^{k-1}) - \exp(q^{k})]},$$ 
for large $c$.
\noindent
\end{proposition} 
\begin{proof}[Proof] 
The log-likelihood function is
$$L(y_1,\dots,y_m; c) = \sum_{j=1}^m \log\left\{(1-q^{y_j})^c - (1-q^{y_j-1})^c\right\}.$$ 
Formally differentiating with respect to $c$, we have the score function as given in \eqref{eq:discrete}.  Squaring and taking
expectations in the case $m=1$, we have Fisher's information per observation:
\begin{displaymath} 
I(c) =  \sum_{y=1}^{\infty} \frac{[\log(1-q^{y})(1-q^{y})^c - \log(1-q^{y-1})(1-q^{y-1})^c]^2}{(1-q^{y})^c - (1-q^{y-1})^c}.
\end{displaymath} 
As $m \to \infty$, from the usual large sample theory of maximum likelihood estimation, $\hat{c}/c$ is asymptotically normally distributed with mean $1$ and variance $1/(m\psi_c)$ where $\psi_c = c^2 I(c)$.
Now let $c \to \infty$ through the sequence $c=q^{-r}$, where $r$ is a positive integer. Writing $y=r+k$, we have  
\begin{displaymath} 
\lim_{c \to \infty} c^2 I(c) = \sum_{k=-\infty}^{\infty} \frac{q^{2k}(q^{-1}-1)^2}{[\exp(q^{k-1}) - \exp(q^{k})]}, 
\end{displaymath}
\noindent
as claimed. 
\end{proof} 

In practice, to solve for $\hat{c}$ in \eqref{eq:discrete}, one iteration of the Newton-Raphson algorithm started from a consistent estimator of $c$ produces an asymptotically efficient estimator \citep{Rao.73}.  A consistent estimator is $\hat{c}=\log (r/m) / \log (1- q^n )$, where $r=| \{y_j; y_j \leq n\} |$ and  $n = \lfloor \log_q(1/2) \rfloor$, if $r \neq 0$, else, set $\hat{c} = T$, the length of the stream observed.   

The statistical efficiency of the maximal-term MLE in the geometric case can be made arbitrarily close to that in the continuous case. For large $c$, the Fisher information is an increasing function of $q$  as $q \to 1$. In particular, for $q = 10/11$, the ARE of the estimator of $c$ based on a sample of maxima from $G_p$ as compared to the estimator based on a random sample of maxima from any continuous distribution is $0.9985$.  For the special case considered by \cite{DF.03} and \cite{FFGM.07} with $p=1/2$, the asymptotic relative efficiency is $0.9304$. 

We note that the estimator $\hat{c}$, based on a sample of maxima from $G_p$, does not have the property of recursive computability, unlike the estimator in the continuous case.  Nevertheless, when $q$ approaches $1$, the geometric distribution is well approximated by the exponential distribution with parameter  $\lambda = - \log
q,$ so the log-likelihood is approximately 
\begin{displaymath}
L(y_1,\dots,y_m ;c) =  m \log (c \lambda) +(c-1) \sum_{j=1}^m \left \{ \log \left (1-e^{-\lambda y_j} \right ) \right \} - \lambda \sum_{j=1}^m y_j.
\end{displaymath} 
For this distribution, the statistic $S_m = \prod_{j=1}^m \left (1 - e^{-\lambda Y_j} \right ) = \prod_{j=1}^m \left (1-q^{Y_j} \right )$ is sufficient for the parameter $c$, and the MLE is $\hat{c} = -m / \log S_m$, so that, to this degree of approximation, recursive estimation is possible.   

\subsection{Storage requirements} \label{sec:storage} 
In this section we determine exponentially decreasing upper bounds on the tail probabilities of our estimators, and show that in the geometric case, the storage requirement of an algorithm implementing the estimation procedure attains the tight lower bound of \cite{IW.03}.
\begin{proposition} \label{prop:tail_bounds}
In the continuous case, the tail error bounds for the estimator $\hat{c}$ given in \eqref{eq:general_max} are
\begin{displaymath}
\P \left (\hat{c} \geq (1+\epsilon) c \right ) \leq \exp(-m \epsilon^2 / C_1), \ \textrm{and} \ \P \left (\hat{c} \leq (1-\epsilon)c \right ) \leq \exp(-m \epsilon^2 / C_2),
\end{displaymath}
where 
\begin{displaymath}
C_1 = \frac{\epsilon^2 (1 + \epsilon)}{-\epsilon + (1 + \epsilon) \log (1 + \epsilon)}, \ C_2 = \frac{\epsilon^2 (1 - \epsilon)}{\epsilon + (1 - \epsilon) \log (1 - \epsilon)}.  
\end{displaymath}
\noindent
In the limit as $\epsilon \to 0$, the constants $C_1$ and $C_2$ tend to 2, so for small $\epsilon$, the tail error bounds are exponentially decreasing in $m \epsilon^2$.    
\end{proposition}
\begin{proof}
In the continuous case, the pivotal quantity $mc/\hat{c}$ has a Gamma distribution with moment generating function $(1-t)^m, t < 1$.   The bounds on the tail probabilities are obtained from the moment generating  function using the method of \cite{Chernoff.52}.  
\end{proof}
In the discrete geometric case, these results hold to arbitrary accuracy by approximating the geometric distribution by an exponential distribution with mean $-\log q$ and $q$ close to 1.  From Proposition~\ref{prop:tail_bounds}, $\hat{c}$ is an $(\epsilon, \delta)$-approximation of $c$ provided that $m = O(\epsilon^{-2})$. The expected value of the maximum order statistic based on a sample of size $c$ from $G_p$ is $O(\log c)$ for fixed $p$ \citep{KP.93}. It follows that the space requirement of an algorithm implementing the estimation procedure in the geometric case is of  order $O \big (\epsilon^{-2} \log(\log c) \big )$, attaining the tight lower bound of \cite{IW.03}.

\section{Random projections} \label{sec:randomproj}
Data sketching via random projections exploits properties of the $\alpha$-stable distribution, introduced by \cite{Levy.24}.  The stability property lies at the heart of the random projection method.  For simplicity, we restrict attention to positive, strictly stable variables of index $\alpha$, for $\alpha \in (0,1)$, having Laplace transform $e^{-\lambda^{\alpha}}$, $\lambda \geq 0$ \citep{Feller.71, Zolotarev.86}.  Let $F_{\alpha}$ denote the distribution function.  The stability property of $F_{\alpha}$ is as follows: if $X_1, X_2 \sim F_{\alpha}$ independently, and $a_1$ and $a_2$ are arbitrary positive constants, then 
\begin{equation} \label{eq:stability}
a_1 X_1 + a_2 X_2 \stackrel{\mathcal{D}}{=} \Big ( a_1^{\alpha} + a_2^{\alpha} \Big )^{1/\alpha} X,
\end{equation} 
where $X \sim F_{\alpha}$ \citep{Feller.71}. 

The random projection method for cardinality estimation proceeds as follows \citep{CDIM.02, Indyk.06}. For $j=1,\ldots,m$ and $\alpha \in (0,1)$ fixed, let $h_j$ be independent hash functions mapping from $\mathcal{I}$ to samples from $F_{\alpha}$, via the usual method of seeding; in practice, this will involve constructing simulated $F_\alpha$ variables from pairs of $\U(0,1)$ variables. Then, update and store the projections $V_j(T)= \sum_{t=1}^T d_t h_j(i_t)$, $j=1,\ldots,m$, to give the data sketch $V_1 ,\ldots,V_m$, where we write $V_j=V_j(T)$ for brevity. By the stability property in \eqref{eq:stability}, we have that   
\begin{equation}\label{eq:scaleparameter}
V_j  = \textstyle{ \sum_{t=1}^T d_t h_j(i_t) = \sum_{i \in \mathcal{I}_T} a_i h_j(i) \stackrel{\mathcal{D}}{=} \ell_\alpha(\mathbf{a}) X_j,}
\end{equation}
where $X_j \sim  F_\alpha$ independently for  $j=1,\dots,m$, and $\ell_\alpha(\mathbf{a}) =  ( \sum_{i \in \mathcal{I}_T} a_i^{\alpha}   )^{1/\alpha}$. In other words, $V_1,\ldots,V_m$ is a sample from a scale family with unknown scale parameter $\ell_\alpha(\mathbf{a})$. It should be noted that when $d_{t} = 1$ for $t = 1,\dots,T$, then $\ell_\alpha(\mathbf{a}) =  ( \sum_{i \in \mathcal{I}_T} n_i^{\alpha}   )^{1/\alpha}$, where $n_i$ is the number of times that item $i$ is observed in the data stream by time $T$.
 
In principle, calculation of the MLE of the scale parameter, $\ell_\alpha(\mathbf{a})$ in \eqref{eq:scaleparameter}, is straightforward. Raising this MLE to the power of $\alpha$ gives the MLE of $\sum_{i \in \mathcal{I}_T} a_i^{\alpha}$, and with $\alpha$ sufficiently small, this produces an approximation to $\hat{c}$. In practice there are severe numerical difficulties in obtaining the MLE when $\alpha$ is small \citep{Nolan.97,Nolan.01}. 

Instead, \cite{CDIM.02} estimate $\ell_\alpha(\mathbf{a})$ by $\tilde{V}/\tilde{\mu}$, where $\tilde{V}$ is the sample median of $\{V_1,\ldots,V_m\}$, and $\tilde{\mu}$ is the numerically determined median of $F_\alpha$. They show that an $(\epsilon,\delta)$-approximation to $c$ can be obtained by choosing $m$ of order $O \left (1/\epsilon^2 \cdot \log(1/\delta) \right )$ and $0<\alpha \leq \epsilon/\log(B)$, where $B$ is an upper bound for the elements of $\mathbf{a}$.   

We adopt a slightly different approach and exploit the limiting distribution of $V_j^\alpha$ for small $\alpha$. 

\begin{proposition}\label{thm:projection}
As $\alpha \to 0$, the random variable
\begin{equation}\label{eq:projection} 
c \sum_{j=1}^m V_j^{-\alpha} \overset{\mathcal{D}}{\to} \G(m,1).
\end{equation}  
Consequently, the variable can be used as an approximate pivot in setting confidence intervals for $c$.  For $\alpha$ small, the estimator $\hat{c} = m / \sum_{j=1}^m V_j^{-\alpha}$ has asymptotic distribution $\N(c,c^2/m)$ as $m \to \infty$. 
\end{proposition}
\begin{proof}
\cite{Zolotarev.86} shows that $X^{\alpha} \overset{\mathcal{D}}{\to} 1/Z$ where $Z \sim \E(1)$, as $\alpha \to 0$. 
It follows from \eqref{eq:scaleparameter} that 
$$ V_j^{\alpha} = \left[\sum_{i \in \mathcal{I}_T} a_i h_j(i)\right]^\alpha \overset{\mathcal{D}}{=} X^\alpha \sum_{i \in \mathcal{I}_T} a_i^\alpha 
\overset{\mathcal{D}}{\to} c/Z,\quad j = 1,\dots, m \quad \text{(independently)},$$ and hence $c \sum_{j=1}^m V_j^{-\alpha} \to \G(m,1)$.
The estimator $\hat{c}$ is obtained by equating the pivot to its mean $m$, and the approximate distribution of $\hat{c}$ then follows from the asymptotic normality of the Gamma distribution.  
\end{proof}
When comparing the estimator $\hat{c}$ above with $\tilde{c} = \left (\tilde{V}/\tilde{\mu} \right )^{\alpha}$ in \cite{CDIM.02}, we are effectively comparing the MLE of the parameter of an exponential distribution with an estimator obtained by equating the sample and population medians.  The ARE of $\tilde{c}$ to $\hat{c}$ is then approximately $48$\% since by using the standard asymptotic distribution of sample medians, we find that $\tilde{c} \sim \N(c, c^2(\log 2)^{-2}/m)$ for large $m$ i.e., $\hat{c}$ is twice as efficient asymptotically as $\tilde{c}$.

At this stage we have shown that the estimators of $c$ using the maximal-term or random projection sketches can have comparable efficiency. This leads us to conjecture that in some sense the methods are essentially equivalent, which we explore in the next section.

\section{Comparison of projection and maximal-term sketches} \label{sec:compare}
In Section~\ref{sec:cont} we show that the efficiency of the maximal-term data sketch does not depend on the particular continuous distribution that is simulated by the hash function.  For the purpose of comparison, we now hash to $F_\alpha$, in both cases.  Note that we are not proposing to use this distribution directly for the maximal term estimator since it has an extremely heavy tail when $\alpha$ is small. Storing the maximum of $c$ such variables, for $c$ large, would require high precision floating point numbers.  

Consider a data stream in the cash register case, observed up to time $T$.  Let $\mathbf{a}$ denote the accumulation vector, and $c$ the cardinality.  For $j=1,\ldots,m$, let $h_j$ be independent hash functions mapping from $\mathcal{I}$ to copies of $X \sim F_{\alpha}$, for fixed $\alpha \in (0,1)$.  Let $\hat{c}_p  = m / \sum_{j=1}^m V_j^{-\alpha}$ be the projection estimator defined in Proposition \ref{thm:projection} and let $\hat{c}_m$ denote the maximal-term estimator in \eqref{eq:general_max} where $M_j = \max_{i \in \mathcal{I}_T} h^\alpha_j(i)$ and $F$ is the distribution function of $X^\alpha$.

\begin{theorem} \label{link_result}
For small $\alpha$, the pivotal quantities for the maximal-term and projection sketches are equivalent, i.e.,
$$ c \left ( \frac{m}{\hat{c}_p} - \frac{m}{\hat{c}_m} \right ) =  c \sum_{j=1}^m V_j^{-\alpha} +  c \sum_{j=1}^m \log F(M_j) \overset{P}{\to} 0, \quad \text{as $\alpha \to 0$},$$
and in particular $V_j^{-\alpha} + \log F(M_j)  \overset{P}{\to} 0$ for each $j=1,\dots,m$.
\end{theorem}
\begin{proof} 
Let $M = \max_{i \in \mathcal{I}_T} X_i^\alpha$ be a typical maximal term in \eqref{eq:general_max} with $X_i \sim F_\alpha, i \in \mathcal{I}_T$ and let $\delta >0 $ be arbitrary.  Since $\P(M < y) \leq \P(X^\alpha < y)$ and $X^{-\alpha} \overset{\mathcal{D}}{\to} \E(1)$ as $\alpha \to 0$ \citep{Zolotarev.86}, there are values $\alpha_0$ and $y_0> 0$ such that $\P(M < y_0) < \delta$ for all $\alpha < \alpha_0$.

Now let $G_\alpha(y)$ be the distribution function of $X^\alpha$, i.e.\ $G_\alpha = F$ in \eqref{eq:general_max}. Since $X^{-\alpha} \overset{\mathcal{D}}{\to} \E(1)$, then $G_\alpha(y) \to \exp(-1/y),$  uniformly in $y>0$,  and consequently $\log G_\alpha(y) \to -1/y$ uniformly in $y > y_0$ as $\alpha \to 0$. It follows, by the usual arguments, that 
\begin{equation}\label{eq:M}
\log G_\alpha(M) + 1/M  \overset{P}{\to} 0, \quad \text{as $\alpha \to 0$}.
\end{equation}

Finally, writing $V = \sum_{i \in \mathcal{I}_T} a_i X_i$ for the typical term in \eqref{eq:scaleparameter}, we have 
$$M^{1/\alpha} a_{\text{min}}  = X_{\text{max}} a_{\text{min}}  \leq V \leq X_{\text{max}} \sum_{i \in \mathcal{I}_T} a_i  = M^{1/\alpha} \sum_{i \in \mathcal{I}_T} a_i,$$
where $X_{\text{max}} = \max_{i \in \mathcal{I}_T} X_i$ and $a_{\text{min}} = \min_{i \in \mathcal{I}_T}a_i$. 
It follows that 
\begin{equation}\label{eq:VM}
a_{\text{min}}^\alpha \leq V^\alpha /M \leq \left ( \sum_{i \in \mathcal{I}_T} a_i \right )^\alpha,
\end{equation} 
and as $\alpha \to 0$, $V^\alpha/M \overset{P}{\to} 1$.  Since both $M$ and $V^{\alpha}$ have proper limiting distributions, this implies that 
$ V^{-\alpha} - M^{-1} \overset{P}{\to} 0 \quad \text{as $\alpha \to 0$},$
and together with \eqref{eq:M} we have
$ \log F(M) + V^{-\alpha} \overset{P}{\to} 0 \quad \text{as $\alpha \to 0$}.$
We have established that the terms in the summations are individually equivalent for small $\alpha$ and since the number of terms, $m$, is finite, the result is proved.
\end{proof}
Note that the specific values of $d_t > 0$ are unimportant in determining the cardinality. For practical purposes, positive values of $d_t$ can be taken to be $1$ and this may have the effect of improving the bounds in \eqref{eq:VM}.
  
\section{Empirical study} \label{sec:simulations}

Table~\ref{table:sim} presents the results of an empirical study comparing various cardinality estimation algorithms on simulated data sets of exact cardinality ranging from $10^4$ to $5\times 10^7$.  We compare the performance of our estimators from Propositions~\ref{thm:maxterm}, \ref{prop:MLE_Geom}, and \ref{thm:projection}, against that of the Hyper-LogLog \citep{FFGM.07}, MinCount \citep{Giroire.09}, and median \citep{CDIM.02} estimators.  We also compute the LogLog estimator \citep{DF.03}, and find that its performance is not comparable; the percent error is consistently above 20$\%$, and as high as 50$\%$ for the low end of cardinalities (results not shown).   Furthermore, we compute the maximal-term estimator with hashing to the positive, $\alpha$-stable distribution ($\alpha = 0.05$); this estimator is compared to the random projection estimator in Theorem~\ref{link_result}.  Again, results are not shown; the positive, $\alpha$-stable distribution, for $\alpha$ close to zero, is very heavy tailed, and numerical difficulties are encountered in estimating the cumulative distribution function in the tail.  Computations are performed on a 64GB supercomputer; the code is written in C, and uses the GSL library, and R (http://www.r-project.org/) packages.  The data is simulated in R.  

\begin{table}[!ht] 
\caption{Comparison of cardinality estimation algorithms on simulated data sets: maximal-term estimators with hashing to the exponential distribution of mean 1, and to the geometric distribution ($\rho = 1.1$), Hyper-LogLog \citep{FFGM.07}, MinCount \citep{Giroire.09}, and random projections estimators (Proposition~\ref{thm:projection} and median estimator of \cite{CDIM.02}) with hashing to the positive, $\alpha$-stable distribution ($\alpha = 0.05$).  The percent error appears in brackets.}\label{table:sim}
\begin{center}
\begin{tabular}{|c|c|c|c|c|c|c|c|}
\hline
$c$ & $m$ & $\hat{c}$ & $\hat{c}$ & Hyper-LogLog & MinCount & $\hat{c}$ & $\tilde{c}$ \\
& & (Prop.~\ref{thm:maxterm}) & (Prop.~\ref{prop:MLE_Geom}) &  &  & (Prop.~\ref{thm:projection}) & (Sec.~\ref{sec:randomproj}) \\
\hline
$10^4$ & $2^9$ & 9543 & 9553 & 8040 & 10261 & 10837 & 10261 \\
& & (4.56) & (4.47) & (19.6) & (2.61) & (8.36) & (2.61) \\
$5\times 10^4$ & $2^9$ & 50190 & 50144 & 43754  & 48594 & 51436 & 51349 \\
& & (0.378) & (0.288) & (12.5) & (2.81) & (2.87) & (2.70) \\
$10^5$ & $2^{10}$ & 102761 & 102916 & 102122 & 98113 & 98527 & 110363\\
& & (2.76) & (2.92) & (2.12) & (1.89) & (1.47) & (10.36) \\
$5 \times 10^5$ & $2^{11}$ & 512056 & 511988 & 431965 & 499698 & 493066 & 502999 \\
& & (2.41) & (2.40) & (13.6) & (0.0602) & (1.39) & (0.600) \\
$10^6$ & $2^{13}$ & 994803 & 994702 & 992408 & 1004610 & 971817 & 1002570 \\
& & (0.520) & (0.530) & (0.759) & (0.461) & (2.82) & (0.257) \\
$5 \times 10^6$ & $2^{14}$ & 5001560 & 4992499 & 4727310 & 5001000 & 4924112 & 5019670 \\
& & (0.0311) & (0.150) & (5.45) & (0.0200) & (1.52) & (0.393) \\
$10^7$ & $2^{14}$ & 9992780 & 9965677 & 9313600 & 10102100 & 9826337 & 9725290 \\
& & (0.0722) & (0.343) & (6.86) & (1.02) & (1.74) & (2.75) \\
$5 \times 10^7$ & $2^{14}$ & 50666000 & 50221623 & 47764000 & 50118300 & 49258239 & 46677800  \\
& & (1.33) & (0.443) & (4.47) & (0.237) & (1.48) & (6.64) \\
\hline
\end{tabular}
\end{center}
\end{table}

Overall, the performance of these algorithms is impressive, particularly on large scales, where a data sketch of size 16384 suffices to estimate cardinality values up to $5 \times 10^7$ with extremely high accuracy.  From the results on asymptotic efficiency, we expect that, with 95$\%$ confidence, our estimates are within 8.66, 6.12, 4.33, 2.17, and 1.53$\%$ for $m \in \{2^9, 2^{10}, 2^{11}, 2^{13}, 2^{14}\}$, respectively, regardless of the size of $c$.  For small scales, the Hyper-LogLog estimator is clearly outperformed by the other estimators.  For the approach based on hashing and storing order statistics, the estimators of Propositions~\ref{thm:maxterm} and \ref{prop:MLE_Geom} have comparable performance to the MinCount estimator.  Similarly, for the approach based on random projections and the stable distribution, the estimator of Proposition~\ref{thm:projection} has comparable performance to the median estimator of \cite{CDIM.02}.  Both in terms of performance, and storage requirements, we prefer the maximal-term estimator of Proposition~\ref{prop:MLE_Geom} with hashing to the Geometric distribution.  

\section{Conclusion} \label{sec:conclude}
In this paper we discuss the problem of cardinality estimation over streaming data, under the assumption that the size of the data precludes the possibility of maintaining a comprehensive list of all distinct data elements observed.  Probabilistic counting algorithms process data elements on the fly in three steps: (i) hash each data element to a copy of a pseudo-random variable, (ii) update a low-dimensional data sketch of the stream, and (iii) discard the data element.   For this purpose, we present two approaches: indirect record keeping using pseudo-random variates and storing either selected order statistics, or random projections.  Both approaches exploit the idea of hashing via the method of seeding.  The data sketch is a random sample of variables whose distribution is parameterised by the cardinality as unknown parameter, and we derive estimators of the cardinality in a conventional statistical framework.   We believe that hashing and data sketching are novel ideas in the statistics literature, and offer great potential for further development in problems of dimension reduction, and online data analysis.

We analyse the statistical properties of our estimators in terms of Fisher information, asymptotic relative efficiency, and error bounds on the estimation error, and the computational properties in terms of recursive computability and storage requirements.  Compared to existing algorithms that employ the same approaches to cardinality estimation, our estimators outperform in terms of ARE with one exception: the probabilistic counting algorithm of \cite{FM.85} that stores a bitmap table, and therefore is far more computationally expensive.  Finally, we demonstrate an unexpected link between the method of maximal-term sketching based on hashing to the $F_\alpha$ distribution, and the method of random projections, showing that the two methods are essentially the same when $\alpha$ is small.  However, since there is no gain in efficiency for the maximal-term sketch in using the $F_\alpha$ distribution, rather than the simpler $\U(0,1)$ distribution, as shown in Section~\ref{sec:cont}, the latter is to be preferred.  Moreover, since we show in Section \ref{sec:discrete} that discrete hash functions are capable of comparable efficiency but with reduced storage requirements, discrete maximal-term methods must be the method of choice.  In fact, algorithms implementing our estimation procedure with discrete maximal-term sketching and geometric hashing attain the tight lower bound on storage requirements for cardinality estimation.   An empirical study estimating cardinalities up to $5 \times 10^{7}$ supports our theoretical results.

\section*{Acknowledgement}
The authors would like to thank Professor Steffen Lauritzen for insightful discussions, and the referees for suggestions that have significantly improved the paper.  Ioana Cosma would like to thank the Department of Statistics at University of Oxford for funding via the Teaching Assistant Bursary scheme.

\bibliographystyle{sjs}
\bibliography{bibliography_SJS}

\begin{thebibliography}{35}
\expandafter\ifx\csname natexlab\endcsname\relax\def\natexlab#1{#1}\fi

\bibitem[{Aggarwal(2007)}]{Aggarwal.07}
Aggarwal, C.~C. (2007).
\newblock \emph{{Data streams: Models and Algorithms.}}
\newblock Springer-Verlag, New York.

\bibitem[{Akella \emph{et~al.}(2003)Akella, Bharambe, Reiter \&
  Seshan}]{ABRS.03}
Akella, A., Bharambe, A., Reiter, M. \& Seshan, S. (2003).
\newblock {Detecting DDoS attacks on ISP networks.}
\newblock In \emph{{ACM SIGMOD/PODS Workshop on Management and Processing of
  Data Streams (MPDS)}}. San Diego, California.

\bibitem[{Alon \emph{et~al.}(1999)Alon, Matias \& Szegedy}]{AMS.99}
Alon, N., Matias, Y. \& Szegedy, M. (1999).
\newblock The space complexity of approximating the frequency moments.
\newblock \emph{J. Comput. System Sci.} \textbf{58}, 137--147.

\bibitem[{Bar~Youssef \emph{et~al.}(2002)Bar~Youssef, Jayram, Kumar, Sivakumar
  \& Trevisan}]{BYJKST.02}
Bar~Youssef, Z., Jayram, T.~S., Kumar, R., Sivakumar, D. \& Trevisan, L.
  (2002).
\newblock Counting distinct elements in a data stream.
\newblock \emph{Lecture Notes in Comput. Sci.} \textbf{2483}, 1--10.

\bibitem[{Carter \& Wegman(1979)}]{CW.79}
Carter, J.~L. \& Wegman, M.~N. (1979).
\newblock Universal classes of hash functions.
\newblock \emph{J. Comput. System Sci.} \textbf{18}, 143--154.

\bibitem[{Chassaing \& Gerin(2006)}]{CG.06}
Chassaing, P. \& Gerin, L. (2006).
\newblock {Efficient estimation of the cardinality of large data sets.}
\newblock In \emph{{Proceedings of 4th Colloquium on Mathematics and Computer
  Science}}, vol. {AG of {\it Discrete Mathematics and Theoretical Computer
  Science Proceedings (DMTCS)}}. pp. 419--422.

\bibitem[{Chen \& Cao(2009)}]{CC.09}
Chen, A. \& Cao, J. (2009).
\newblock {Distinct counting with a self-learning bitmap.}
\newblock In \emph{{Proceedings of IEEE International Conference on Data
  Engineering}}. pp. 1171--1174.

\bibitem[{Chernoff(1952)}]{Chernoff.52}
Chernoff, H. (1952).
\newblock A measure of asymptotic efficiency for tests of a hypothesis based on
  the sum of observations.
\newblock \emph{Annals of Mathematical Statistics} \textbf{23}, 493--507.

\bibitem[{Cormode \emph{et~al.}(2003)Cormode, Datar, Indyk \&
  Muthukrishnan}]{CDIM.02}
Cormode, G., Datar, M., Indyk, P. \& Muthukrishnan, S. (2003).
\newblock {Comparing data streams using Hamming norms (how to zero in).}
\newblock \emph{{IEEE Transactions on Knowledge and Data Engineering}}
  \textbf{15}, 529--540.

\bibitem[{Cormode \& Muthukrishnan(2005{\natexlab{a}})}]{CM.04a}
Cormode, G. \& Muthukrishnan, S. (2005{\natexlab{a}}).
\newblock {An improved data stream summary: the Count-Min sketch and its
  applications.}
\newblock \emph{J. Algorithms} \textbf{55}, 58--75.

\bibitem[{Cormode \& Muthukrishnan(2005{\natexlab{b}})}]{CM.04b}
Cormode, G. \& Muthukrishnan, S. (2005{\natexlab{b}}).
\newblock What's new: Finding significant differences in network data streams.
\newblock \emph{{IEEE/ACM Transactions on Networking (TON)}} \textbf{13},
  1219--1232.

\bibitem[{Durand \& Flajolet(2003)}]{DF.03}
Durand, M. \& Flajolet, P. (2003).
\newblock Loglog counting of large cardinalities.
\newblock \emph{Lecture Notes in Comput. Sci.} \textbf{2832}, 605--617.

\bibitem[{Feller(1971)}]{Feller.71}
Feller, W. (1971).
\newblock \emph{An introduction to probability theory and its applications},
  vol.~2.
\newblock John Wiley \& Sons, Inc., New York, 2nd edn.

\bibitem[{Fisher(1925)}]{Fisher.25}
Fisher, R.~A. (1925).
\newblock {Theory of statistical estimation.}
\newblock \emph{Math. Proc. Cambridge Philos. Soc.} \textbf{22}, 700--725.

\bibitem[{Flajolet \emph{et~al.}(2007)Flajolet, Fusy, Gandouet \&
  Meunier}]{FFGM.07}
Flajolet, P., Fusy, E., Gandouet, O. \& Meunier, F. (2007).
\newblock {HyperLogLog: the analysis of a near-optimal cardinality estimation
  algorithm.}
\newblock In \emph{{Proceedings of Conference on Analysis of Algorithms}}, vol.
  {AH of {\it Discrete Mathematics and Theoretical Computer Science Proceedings
  (DMTCS)}}. pp. 127--146.

\bibitem[{Flajolet \& Martin(1985)}]{FM.85}
Flajolet, P. \& Martin, G.~N. (1985).
\newblock Probabilistic counting algorithms for database applications.
\newblock \emph{J. Comput. System Sci.} \textbf{31}, 182--209.

\bibitem[{Giroire(2009)}]{Giroire.09}
Giroire, F. (2009).
\newblock Order statistics and estimating cardinalities of massive data sets.
\newblock \emph{Discrete Appl. Math.} \textbf{157}, 406--427.

\bibitem[{Hammersley(1950)}]{Hammersley.50}
Hammersley, J.~M. (1950).
\newblock On estimating restricted parameters.
\newblock \emph{J. Roy. Statist. Soc. Ser. B} \textbf{12}, 192--240.

\bibitem[{Harvey \emph{et~al.}(2008)Harvey, Nelson \& Onak}]{HNO.08.2}
Harvey, N. J.~A., Nelson, J. \& Onak, K. (2008).
\newblock {Sketching and streaming entropy via approximation theory.}
\newblock In \emph{{49th Annual IEEE Symposium on Foundations of Computer
  Science (FOCS)}}. pp. 489--498.

\bibitem[{Indyk(2006)}]{Indyk.06}
Indyk, P. (2006).
\newblock Stable distributions, pseudorandom generators, embeddings, and data
  stream computation.
\newblock \emph{J. ACM} \textbf{53}, 307--323.

\bibitem[{Indyk \& Woodruff(2003)}]{IW.03}
Indyk, P. \& Woodruff, D.~P. (2003).
\newblock Tight lower bounds for the distinct elements problem.
\newblock In \emph{{Annual Symposium on Foundations of Computer Science
  (FOCS)}}, vol.~44. Cambridge, MA, USA, pp. 283--289.

\bibitem[{Kane \emph{et~al.}(2010)Kane, Nelson \& Woodruff}]{KNW.10}
Kane, D.~M., Nelson, J. \& Woodruff, D.~P. (2010).
\newblock {An optimal algorithm for the distinct elements problem.}
\newblock In \emph{{Proceedings of the 29th Symposium on Principles of Database
  Systems (PODS)}}. Indiana, USA, pp. 41--52.

\bibitem[{Kirschenhofer \& Prodinger(1993)}]{KP.93}
Kirschenhofer, P. \& Prodinger, H. (1993).
\newblock A result in order statistics related to probabilistic counting.
\newblock \emph{Computing} \textbf{51}, 15--27.

\bibitem[{Knuth(1998)}]{Knuth.98}
Knuth, D.~E. (1998).
\newblock \emph{The art of computer programming: Sorting and searching},
  vol.~3.
\newblock Addison-Wesley, Massachusetts, 2nd edn.

\bibitem[{Lauritzen(1988)}]{Lauritzen.88}
Lauritzen, S.~L. (1988).
\newblock Extremal families and systems of sufficient statistics.
\newblock In \emph{Lecture notes in statist.}, vol.~49. Springer, New York.

\bibitem[{Lehmann \& Scheff\'e(1950)}]{LS.50}
Lehmann, E.~L. \& Scheff\'e, H. (1950).
\newblock {Completeness, similar regions and unbiased estimation. Part I.}
\newblock \emph{Sankhy$\bar{a}$} \textbf{10}, 305--340.

\bibitem[{L\'evy(1924)}]{Levy.24}
L\'evy, P. (1924).
\newblock {Th\'eorie des erreurs. La loi de Gauss et les lois exceptionelles.}
\newblock \emph{Bull. Soc. Math. France} \textbf{52}, 49--85.

\bibitem[{Muthukrishnan(2005)}]{Muthukrishnan.05}
Muthukrishnan, S. (2005).
\newblock \emph{Data streams: Algorithms and applications}.
\newblock Now Publishers Inc, Cambridge, Massachusetts, 1st edn.

\bibitem[{Nisan(1992)}]{Nisan.92}
Nisan, N. (1992).
\newblock {Pseudorandom generators for space-bounded computation.}
\newblock \emph{Combinatorica} \textbf{12}, 449--461.

\bibitem[{Nolan(1997)}]{Nolan.97}
Nolan, J.~P. (1997).
\newblock Numerical calculation of stable densities and distribution functions.
\newblock \emph{Stoch. Models} \textbf{13}, 759--774.

\bibitem[{Nolan(2001)}]{Nolan.01}
Nolan, J.~P. (2001).
\newblock Maximum likelihood estimation and diagnostics for stable
  distributions.
\newblock In O.~E. Barndorff-Nielsen, T.~Mikosch \& S.~I. Resnick, eds.,
  \emph{L\'evy processes: Theory and applications}. Birkh{\"a}user, Boston, pp.
  379--400.

\bibitem[{Press \emph{et~al.}(2007)Press, Teukolsky, Vetterling \&
  Flannery}]{PTVF.07}
Press, W.~H., Teukolsky, S.~A., Vetterling, W.~T. \& Flannery, B.~P. (2007).
\newblock \emph{Numerical recipes: The art of scientific computing}.
\newblock Cambridge University Press, New York, NY.

\bibitem[{Rao(1973)}]{Rao.73}
Rao, C.~R. (1973).
\newblock \emph{Linear statistical inference and its applications}.
\newblock John Wiley \& Sons, Inc., New York, 2nd edn.

\bibitem[{Whang \emph{et~al.}(1990)Whang, Vander-Zanden \& Taylor}]{WVT.90}
Whang, K.-Y., Vander-Zanden, B.~T. \& Taylor, H.~M. (1990).
\newblock A linear-time probabilistic counting algorithm for database
  applications.
\newblock \emph{ACM Transactions on Database Systems} \textbf{15}, 208--229.

\bibitem[{Zolotarev(1986)}]{Zolotarev.86}
Zolotarev, V.~M. (1986).
\newblock \emph{One-dimensional stable distributions}.
\newblock American Mathematical Society, Providence, RI.

\end{thebibliography}

\vspace{0.5cm}

\noindent
Ioana A. Cosma, Statistical Laboratory, Centre for Mathematical Sciences, University of Cambridge, Wilberforce Road, Cambridge, CB3 0WB, United Kingdom.\\
E-mail: ioana@statslab.cam.ac.uk

\end{document}